\documentclass[aps]{revtex4}
\usepackage{graphicx}
\begin{document}
\title{Extinction Rates for Fluctuation-Induced Metastabilities : A Real-Space WKB Approach}
\author{David A. Kessler}
\affiliation{Dept. of Physics, Bar-Ilan University, Ramat-Gan 52900
Israel}
\author{Nadav M. Shnerb}
\affiliation{Dept. of Physics, Bar-Ilan University, Ramat-Gan 52900
Israel}
\begin{abstract}
The extinction of a single species due to demographic
stochasticity is analyzed. The discrete nature of the individual
agents and the Poissonian noise related to the birth-death processes
result in local extinction of a metastable population, as the system
hits the absorbing state. The Fokker-Planck formulation of that
problem fails to capture the statistics of large deviations from the
metastable state, while approximations appropriate close to the absorbing state
become, in general,  invalid as the population becomes large. To
connect these two regimes, a master equation based on a real space WKB
method is presented, and is shown to yield an excellent
approximation for the decay rate and the extreme events statistics
all the way down to the absorbing state. The details of the
underlying microscopic process, smeared out in a mean field
treatment, are shown to be crucial for an exact determination of the
extinction exponent. This general scheme is shown to reproduce the
known results in the field, to yield new corollaries and to fit
quite precisely the numerical solutions. Moreover it allows for
systematic improvement via a series expansion where the small parameter is the
inverse of the number of individuals in the metastable state. 
\end{abstract}
\maketitle

Local extinction due to demographic stochasticity is a key issue in
the analysis of persistence and viability of small populations.\cite{0,1,2} 
In particular, it allows ecologists to identify
endangered species and to specify conservation policies, it dictates
the appearance and disappearance of favored and neutral genetic
mutations \cite{5} and is of importance in the determination of the
critical population size needed to support an epidemic.\cite{3} As
in most cases of rare and extreme events  the  important quantities
to be measured and compared with the theory are  the expected
extinction time of the populations,\cite{1} and the probability
distribution close to the absorbing state. Technically, however, the
effect of demographic stochastisity has traditionally been taken into account using
some version of a Fokker-Planck equation, where the "diffusion
coefficient" is a function of the population size.\cite{1}

Recently, much interest has been focussed on the calculation of
extinction rates for systems whose macroscopic dynamics exhibits a
stable state which is nevertheless only metastable due to rare
fluctuations which can drive the system to extinction.\cite{ek,b,c,am}
 In particular, it has been realized that in
general the Fokker-Planck (FP) expansion about the (meta-)stable
state is incapable of predicting the extinction rate. This is due to
the fact that the Fokker-Planck expansion is only valid for up to
 $O(N^{2/3})$ fluctuations to the large, $O(N)$, number of
particles in the metastable state. The FP approximation  fails to
correctly describe the very large fluctuations necessary to reach
the absorbing state of zero particles. The FP treatment also smears
out the microscopic differences between processes as it reflects a
local analysis close to the metastable fixed point. In order to get
the correct  statistics for rare and extreme events one should base the
estimate on the exact Master equation that describes the stochastic
process, and to employ the method of extreme statistics, or more
simply put, the WKB approximation, to solve the relevant master
equation.

Elgart and  Kamenev \cite{ek} made an interesting observation in
this context: using the Peliti-Doi \cite{pd,cardy} technique to map
the exact master equation into a "quantum mechanical" problem
(Schroedinger-like equation in imaginary time with second
quantized Hamiltonian) they were able to identify the classical
trajectory that connects the metastable fixed point and the
absorbing state. This identification allows them to calculate the
classical  ("geometrical optics") action along this trajectory, a
first approximation to the extinction time. Assaf and Meerson\cite{am} then suggest a general spectral method to improve beyond the
Elgart-Kamenev results, employing the generating function formalism
and using the Sturm-Liouville theory of linear differential
operators.

In this paper, we will present a general scheme to deal with the
local extinction problem, based on the time-independent "real space" WKB
approximation (unlike Refs. \cite{ek, am} who used a time-dependent
momentum space presentation). The method presented is easy to use,
its intuitive meaning is transparent, and its range of applicability
covers, essentially, any single species problem.

This paper will be organized as follows: In the next section we
exemplify the technique for what is perhaps {\em the} archetypical
problem in this class, a logistic birth-death process of a single
species. Beside its importance, the solution of this example demands
the use of all the components of the technique - a Fokker-Planck
solution applicable close to the metastable fixed point, small $n$
approximation close to the absorbing state  and a WKB solution that encompasses the FP regime and connects to the small $n$ region.
 This model, thus,  serves also  as a nice
pedagogical introduction. The third section deals with a similar
birth death process, but when the number of offsprings at each birth
event is two, as in the case of domain walls in magnetic systems.
Here, a series of mathematical "miracles" occur, which allow for a
simple calculation of the extinction rate (for the case of an
initial even number of particles) without recourse to the WKB
method. In the fourth section the effect of a single agent death
term is incorporated, and in the last section the marginal case of
neutral mutation is analyzed. We then conclude with a summary and
some final observations.

\section{Stochasticity, logistic growth and extinction}

In this section we study the stochastic dynamics of a combination of
two fundamental processes: particles giving birth to new particles
at rate $\alpha$; and pair annihilation at rate $\beta$. In that
case  the average number of particles is about $\alpha/\beta$. The
full "physical optics" solution for both the probability
distribution, $P_n$, of the metastable state and the extinction
rate, in the limit where $\beta \ll \alpha$ is given, based on a WKB
approximation for the exact master equation.   We confirm our
calculations by comparison to a direct numerical solution of the
master equation.

The microscopic rules that govern this process are:
\begin{eqnarray}
P&\stackrel{\alpha}{\to}& 2P\nonumber\\
P+P&\stackrel{\beta}{\to}& 0
\label{basic}
\end{eqnarray}
The exact master equation for $P_n$, the probability of having $n$
particles, is
\begin{equation}\label{master}
\dot{P}_n = \alpha\left[-nP_n + (n-1)P_{n-1}\right] +
\frac{\beta}{2}\left[-n(n-1)P_n + (n+2)(n+1)P_{n+2}\right]
\end{equation}
At the mean-field level the process is described by the reaction
equation,
\begin{equation}
\dot{n}=\alpha n - \beta n(n-1)
\end{equation}
which has the stable solution $n=\alpha/\beta + 1$. Technically,
this expression may be derived from the exact master equation
(\ref{master}) by calculating the time derivative of the average
population $\dot{\langle n \rangle} \equiv \sum n \dot{P_n}$ and
using the approximation $\langle n^2 \rangle = \langle n \rangle^2$.
The stable solution becomes metastable due to the effect of
stochastisity  (in particular, all particles may annihilate each
other and the system will be stuck in the absorbing state $P_0 =1$).
Our aim is to calculate the typical time of this extinction event.

Since that stochastic process has no memory (a Markov process) it
may be described by a transition matrix that specify the rates to
pass from one microscopic configuration to the other. Clearly, the
absorbing  state $\{P_0 =1, \ P_n =0 \ \forall n \neq 0 \}$ is an
eigenvector of that matrix with an eigenvalue 0. All other
eigenstates admit negative eigenvalues, and we denote the absolute
value of the  highest of these eigenvalues as $\Gamma$. The
corresponding eigenvector is the stochastic metastable state, so our
mission is to calculate $\Gamma$. Our main interest is  in the case
$\alpha/\beta \gg 1$ so that the typical number of particles is
large.  This implies that   the probability to reach the absorbing
state is, as we will calculate, exponentially small.

The metastability of the system implies that at long times the $P_n$
decay exponentially as $e^{-\Gamma t}$.  Thus we need to solve the
master equation with the left-hand side replace by $-\Gamma P_n$.
However, since $\Gamma$ is exponentially small, we can drop this
term altogether. Technically this implies that we only have to solve for
a steady state vector rather than doing time dependent semiclassical
analysis which is much more complicated.

\subsection{Fokker-Planck equation and its limitations}

The standard approach for solving the now time-independent master
equation is to transform it into a Fokker-Planck (FP) equation
\cite{0,2}. The nominal prescription for doing this is to expand
$P_{n \pm 1}$, etc. in a Taylor series, dropping terms involving
more than two derivative.  However, phrasing the problem this way
does not explain why, or more specifically, when this procedure is
justified; moreover,  the emerging equation is not unique - the same
expansion may be done for $n P_n$, for example, yielding different
equation. The real justification underlying the Fokker-Planck
approximation is that, for small $\beta/\alpha$, $P_n$ is a smooth
function of the $O(1)$ variable
\begin{equation}
 y\equiv \left(\frac{\beta}{\alpha}\right)^{1/2}\left(n-\frac{\alpha}{\beta}\right)
 \end{equation}
 Then, $P_{n\pm 1}$ is equal to $P(y\pm\sqrt{\beta/\alpha})$ and so may be expanded formally with regard to the small parameter
 $\beta / \alpha$. The resulting series may be written as:
 \begin{equation} \label{pert}
 \alpha \left[ \hat{{\cal L}}_0 P(y) + \sqrt{\frac{\beta}{  \alpha}} \  \hat{{\cal L}}_1 P(y) + {\cal O} \left( \frac{\beta}{  \alpha}\right) +  ... \right] = 0
 \end{equation}
 where
\begin{equation}
 \hat{{\cal L}}_0 = \left(\frac{d}{dy}(yP) + \frac{3}{2}\frac{d^2}{dy^2}P\right)
 \end{equation}
 and
\begin{equation}
 \hat{{\cal L}}_1 = \frac{1}{2} \left(P''' + 5yP'' + 8P' + 4yP + 2y^2 P \right)
 \end{equation}
This perturbative expansion is justified (at least in the sense of asymptotic series) if, expanding $P(y)=P_0(y) + \sqrt{\beta}{\alpha} P_1(y) + \ldots$, the
correction term $\sqrt{\beta/\alpha} P_1(y) \ll P_0(y)$. Plugging this
series into (\ref{pert}) and collecting terms order by order one
finds that
\begin{equation}
P_0(y)=Ce^{-y^2/3}=Ce^{-\beta (n - \alpha/\beta)^2/3\alpha}.
\end{equation}
Then, looking at the equation for $P_1$
\begin{equation}
 \hat{{\cal L}}_0 P_1(y)  = -\sqrt{\frac{\beta}{  \alpha}} \  \hat{{\cal L}}_1
 P_0(y),
 \end{equation}
one is able to identify that the leading correction is proportional
to $ y^3 \sqrt{\beta/\alpha}P_0(y)$.  Thus the
the FP equation is only valid up to $(n- \alpha/\beta ) \sim
(\alpha/\beta)^{2/3}$, as mentioned above. This limit on the FP
reliability is clearly demonstrated in Figure \ref{nppfig}, where
for a metastable population of 100 individuals  the FP solution is a
good approximation to the exact probability distribution  between
100 and 70.

To find the decay rate, however, we need to have a solution valid
down to $n=1$, and the FP solution does not suffice. One can try to
consider the low $n$ limit of the master equation.

\subsection{Probability distribution close to extinction}

\begin{figure}[h]
\center{\includegraphics[width=0.8\textwidth]{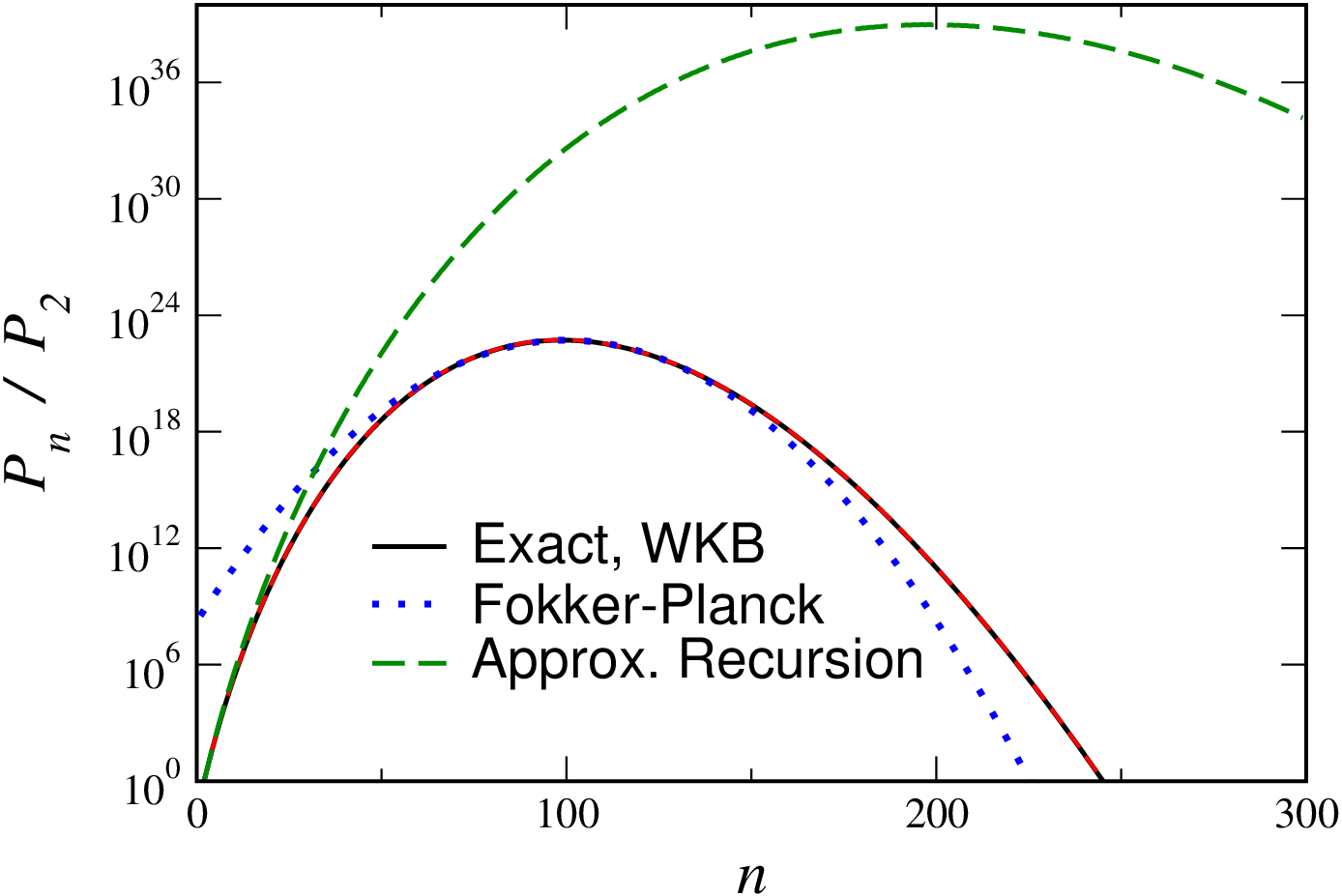}}
\caption{Probability distribution $P_n/P_2$ for $\beta=0.01$, $\alpha=1$, for the
basic model, Eq. (\ref{basic}), together with the WKB
approximation, Eq. (\ref{wkbq}), together with (\ref{wkbs}),  (\ref{npqfin}), and
(\ref{npa}); the FP approximation, Eq. (\ref{npfp}); and the low/intermediate-$n$ result,
Eq. (\ref{nplown}). Note that on the scale of the figure, the WKB approximation is
indistinguishable from the exact result.}
\label{nppfig}
\end{figure}

In the vicinity of the absorbing state one may use a simplified form of the master equation,  exploiting the fact that $P_n$ is a rapidly growing
function of $n$. Then, $P_n \ll P_{n+2}$ and $P_{n-1} \ll P_n$,
yielding the simplified recursion relation:
\begin{equation}
P_{n+2}=\left(\frac{2\alpha}{\beta}\right)\frac{n}{(n+2)(n+1)}P_n
\label{nplownrec}
\end{equation}
This approximate recursion relation leaves the even and odd $n$'s decoupled, and has
the solution
\begin{eqnarray}
P_{2k}&=&\left(\frac{4\alpha}{\beta}\right)^{k-1}\frac{2(k-1)!}{(2k)!}P_2\nonumber\\
P_{2k+1}&=&\left(\frac{\alpha}{\beta}\right)^k\frac{1}{(2k+1)k!}P_1
\label{nplown}
\end{eqnarray}
with $P_1$, $P_2$ arbitrary.
Examining Eq. (\ref{nplownrec}), we see that if $P_1 \ll P_2$, then as long as $n \ll \alpha/\beta$, our
assumption leading to the approximate recursion
relation is valid.
It is also clear that the rapid rise of the $P_n$'s slows as $n$ increases, with
the $P_n$'s reaching
a maximum at $n=2\alpha/\beta$.  Clearly, however, the maximum probability state is at $n\approx\alpha/\beta$, so the recursion relation must fail
before the bulk regime. In fact,
our recursion relation works up to $\beta n/\alpha\ll 1$ whereas the FP solution only works when $\beta n/\alpha \approx 1$.  There is no way to directly connect these
two regimes.  This is seen clearly in Fig. \ref{nppfig}, where the recursion relation
and FP solutions are shown for the case $\beta/\alpha=0.01$. The resolution to this
problem lies in the WKB method, which will allow us to connect the
recursion relation results to the FP regime.

\subsection{WKB approximation and the extinction rate - the leading term}

To do WKB for our difference equation,\cite{bender} we write $P_n=e^{S_n}$, where $S_n$ is assumed to be a
smooth function of $n$, so that $S_{n\pm1}\approx S_n \pm S'_n$.  Since we
already know (from the FP treatment) that the probability profile in
the bulk is a Gaussian with width proportional to the square root of
the metastable population, the quality of this approximation is
controlled. Writing $\Lambda_n\equiv e^{S_n'}$, we have, assuming
$n\gg 1$,
\begin{equation}
0=\alpha\left(-n + \frac{n}{\Lambda_n}\right)+\frac{\beta}{2}\left(-n^2 + n^2\Lambda_n^2\right)
\end{equation}
or, simplifying,
\begin{eqnarray}
0&=&\beta n \Lambda_n^3 - (2\alpha + \beta n)\Lambda_n + 2\alpha\nonumber\\
&=&(\Lambda_n-1)(\beta n \Lambda_n^2 + \beta n\Lambda_n - 2\alpha)
\label{lambdan}
\end{eqnarray}
where we have factored out the trivial $\Lambda_n=1$ root which is a result of conservation
of probability.  The other two roots for $\Lambda$ are
\begin{equation}
\Lambda_n=-\frac{1}{2} \pm \frac{\sqrt{1 + \frac{8\alpha}{\beta n}}}{2}\ ,
\end{equation}
of which the larger, positive, one is relevant for us, since we want $P_n$ to be an increasing function.  This implies that
\begin{equation}
S'_n=\ln\left(\frac{\sqrt{1+\frac{8\alpha}{\beta n}}-1}{2}\right)
\end{equation}
Integrating, we find that
\begin{equation}
S_n=S_0+n\ln\left(\sqrt{1+\frac{8\alpha}{\beta n}}-1\right) + \frac{1}{2}n\sqrt{1+\frac{8\alpha}{\beta n}} - \frac{2\alpha}{\beta}\ln\left(\frac{\beta n}{4\alpha} +1+ \frac{\beta n}{4\alpha}\sqrt{1 + \frac{8\alpha}{\beta n}}\right) - n\ln(2) + \frac{1}{2}n
\label{wkbs}
\end{equation}
The first important point to notice is that $\alpha/(\beta n)$
extrapolates, in the interesting region, from $\infty$ as $n \to 0$
to unity, where $n \to \alpha/\beta$. In the first case $S'$ scales
logarithmically with $\alpha/(\beta n)$, hence $S_n$ is proportional
to $\alpha/(\beta n)$ and is large. In the second regime $S'$ is
almost constant and $S_n$ scale with $n$ which is also, in that
case, large. Thus  all the way to extinction $S_n$  is large, as
expected. Evaluating $\Delta S\equiv S_{\alpha/\beta}-S_0$, one
obtains
\begin{equation}
\Delta S =\frac{2\alpha}{\beta}\left(1-\ln(2)\right)
\end{equation}
in agreement with the result of Elgart and Kamenev.\cite{ek}

We can make the connection to the formalism presented in \cite{ek}
even more explicit, if we express the relations in terms of
$n(\Lambda)$ as opposed to $\Lambda_n$. From Eq. (\ref{lambdan}),
\begin{equation}
n(\Lambda)=\frac{2\alpha}{\beta\Lambda(\Lambda+1)}
\end{equation}
thus
\begin{eqnarray}
\Delta S&=&\int_0^{\alpha/\beta} \ln(\Lambda_n) dn\nonumber\\
&=& -\int_1^\infty \ln(\Lambda) \frac{dn}{d\Lambda} d\Lambda\nonumber\\
&=& -\ln(\Lambda)n(\Lambda)|_1^\infty + \int_1^\infty \frac{n(\Lambda)}{\Lambda} d\Lambda \nonumber\\
&=& \int_1^\infty \frac{n(\Lambda)}{\Lambda} d\Lambda.
\end{eqnarray}
If we now introduce the "momentum"  $p\equiv1/\Lambda$ and the
"coordinate" $q\equiv n\Lambda$, the expression for $\Delta S$ may
be rewritten as
\begin{equation}
\Delta S = \int_0^1 q dp
\end{equation}
where
\begin{equation}
q(p)=\frac{2\alpha}{\beta}\frac{p}{1+p}
\end{equation}
precisely reproducing Elgart-Kamenev equations for the action and
the semiclassical escape path.  The physical meaning of the
"momentum" $p$ is now clarified: it is the  inverse of the
geometrical growth rate of the quasistatic probability distribution.

We now need to confirm that there exists an overlap region between the WKB solution and the  $n\ll \alpha/\beta$ recursion regime.   For $n\ll
\alpha/\beta$ our WKB solution for $S_n$ may be approximated by,
\begin{equation}
S_n \approx S_0 + \frac{1}{2}n\left(1+\ln\left(\frac{2\alpha}{\beta n}\right)\right)
\end{equation}
To compare this with the recursion relation results, Eq.
(\ref{nplown}), in the limit $n\gg 1$:
\begin{eqnarray}
\ln(P_n^{\it even})&\approx& \frac{1}{2}n\left(1+\ln\left(\frac{2\alpha}{\beta n}\right)\right) +
\frac{3}{2}\ln(2)-\ln\left(\frac{4\alpha n}{\beta}\right)+\ln(P_2)\nonumber\\
\ln(P_n^{\it odd}) &\approx& \frac{1}{2}n\left(1+\ln\left(\frac{2\alpha}{\beta n}\right)\right) -\frac{1}{2}\ln\left(\frac{2\pi\alpha n^2}{\beta}\right) + \ln(P_1)
\end{eqnarray}
Indeed, the leading order asymptotics agrees.

In the other extreme  our WKB result  coincides with  the FP
treatment. Expanding the WKB solution for small $y\equiv
\sqrt{\beta/\alpha}(n-\alpha/\beta)$ yields
\begin{equation}
\ln(S_n)=\ln(\Delta S - y^2/3)
\end{equation}
indeed reproducing the FP solution.

\subsection{WKB approximation and the extinction rate - first order corrections}

 The explicit real space, time-independent WKB analysis allows one to go beyond the
 Elgart-Kamenev "geometrical optics" results and to obtain the leading corrections.
 In the previous subsection  the generic substitution  $P_n =
 \exp(S_n)$ was implemented
 and is justified {\em ex post facto} by the fact that $S_n$ turns out to be ${\cal O} (\alpha / \beta) >> 1$. To proceed let us assume that
 \begin{equation}\label{zz}
 P_n =
 \exp(S_0(n)+S_1(n)+S_2(n)+...)
 \end{equation}
 where $S_0(n)$ is the leading order $S_n$ found above and $S_m(n)$ is assumed to be  $O((\beta /
 \alpha)^{m-1})$. Beginning with the growth part of the master
 equation $T_\alpha\equiv\alpha [ -n P_n + (n-1) P_{n-1}]$, plugging in Eq. (\ref{zz}) and
 expanding the small (${\cal O}  (\beta / \alpha)$) terms in the
 exponent one has  (note that any derivative adds a $(\beta /
 \alpha)$ factor):
\begin{eqnarray}\label{zz2}
T_\alpha&=& \alpha e^{S_0(n)+S_1(n)+S_2(n)} \left( -n  + (n-1)
 e^{ - S_0^{'}(n)} \left[ 1 + \frac{1}{2} S_0^{''}
 - S_1^{'} \right]
 \right) \nonumber \\ &=& \alpha n e^{S_0 + S_1+S_2} \left[
 \left ( -1+\frac{1}{\Lambda_n}\right ) + \frac{1}{\Lambda_n} \left( \frac{
 S_0^{''}}{2} - S' - \frac{1}{n} \right) \right]
 \end{eqnarray}
 While the first, ${\cal O}(1)$, term in the bracket was used to determine
 $\Lambda$, the second ${\cal O}(\beta / \alpha)$  term will be used here in order to find the
 function $S_1$. Repeating that procedure and collecting the leading
 corrections from the annihilation part of the Master equation one
 finds,
 \begin{equation}
 0=\frac{\alpha}{\Lambda_n}\left(-S_1' + \frac{S_0''}{2} - \frac{1}{n}\right) + \frac{\beta n}{2} \left(\frac{1}{n} + 2\Lambda_n^2S_1' + 2\Lambda_n^2S_0'' + \frac{3}{n}\Lambda_n^2\right)
 \end{equation}
 Things simplify if we write $S_1$ as a function of $\Lambda$:
 \begin{equation}
\left [\frac{\alpha}{\Lambda}-\beta n\Lambda^2\right]\frac{d}{d\Lambda}S_1=\frac{1}{2\Lambda}\left[\frac{\alpha}{\Lambda}+2\beta n\Lambda^2\right] + \frac{n'(\Lambda)}{2n}\left[-\frac{2\alpha}{\Lambda}+\beta n + 3\beta n\Lambda^2 \right]
 \end{equation}
so that all the coefficient functions are rational functions of $\Lambda$.
The solution of this equation is best expressed in terms of $Q_n\equiv e^{S_1(n)}$:
\begin{equation}
 Q_n=A\frac{\sqrt{\Lambda_n}(\Lambda_n+1)^2}{\sqrt{2\Lambda_n+1}}
 \label{npqfin}
 \end{equation}
 so that to this order
 \begin{equation}
 P_n\approx Q_n e^{S_n}
 \label{wkbq}
 \end{equation}
We see that $Q_n$ diverges as $n^{-1}$ as $n \to 0$, since $\Lambda_n$ diverges there, and vanishes as $n^{-1/2}$ for large $n$, where $\Lambda$ vanishes.
Interestingly enough, there is no turning point, and the WKB approximation is good
everywhere.  This holds despite the fact that the coefficient of $Q_n'$ vanishes at
$\Lambda=1$, where $S'$ vanishes, as is typical for WKB problems.  In our case,
the right hand side also vanishes at $\Lambda=1$, so $Q$ is regular there.  We suspect
that this is a consequence of the effective vanishing of $\Gamma$, so that the FP equation admits a trivial first integral, and so is effectively of first order.

\begin{figure}[h]
\center{\includegraphics[width=0.7\textwidth]{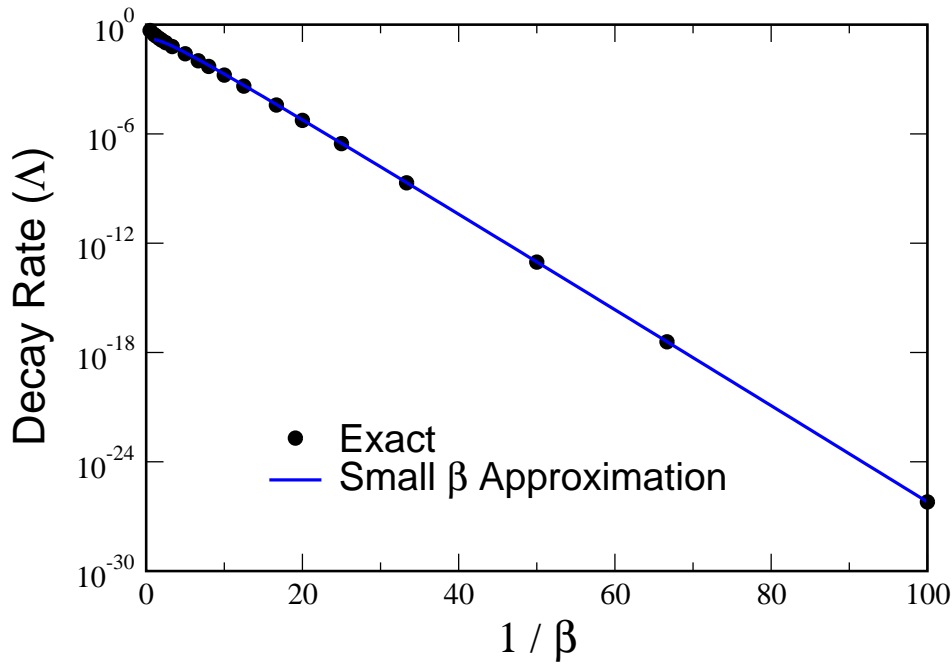}}
\caption{Decay rate vs. $1/\beta$, for the basic model, Eq. (\ref{basic}), with $\alpha=1$, together
with the analytic approximation, Eq. (\ref{npfinal}).  Here and in all other figures, the numerical calculation of the decay rate was performed using a quadruple-precision version of the sparse matrix eigenvalue solver ARPACK,\protect{\cite{arpack}} applied to the master equation transition matrix, with the absorbing state eliminated.}
\label{np}
\end{figure}

\begin{figure}[h]
\center{\includegraphics[width=0.7\textwidth]{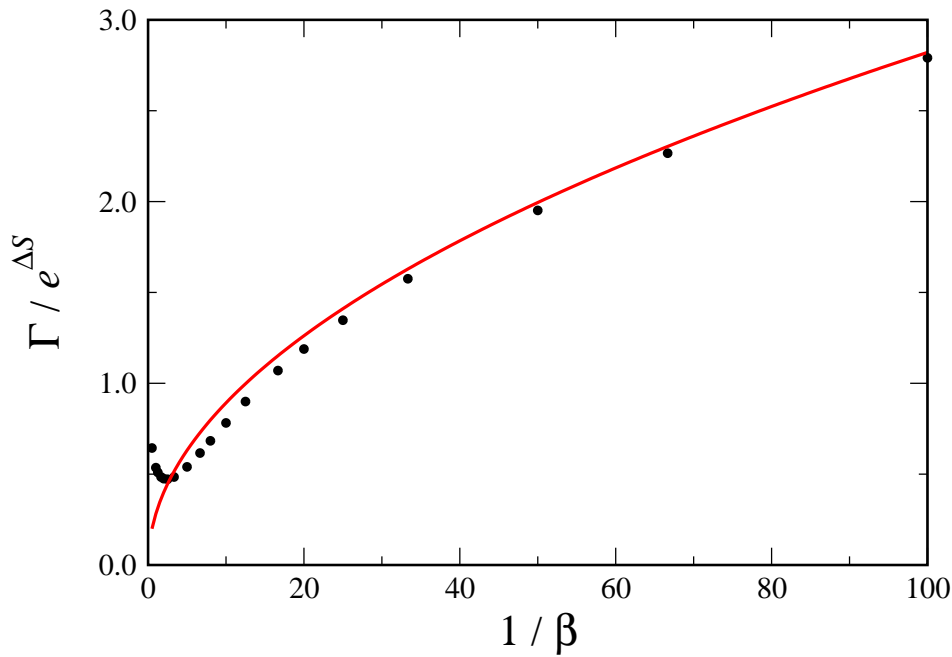}}
\caption{Ratio of $\Gamma$ to $e^{\Delta S}$, for the basic model, Eq. (\ref{basic}), compared to the WKB prediction,
as a function of $\beta$, for $\alpha=1$.}
\label{pref}
\end{figure}

\begin{figure}[h]
\center{\includegraphics[width=0.7\textwidth]{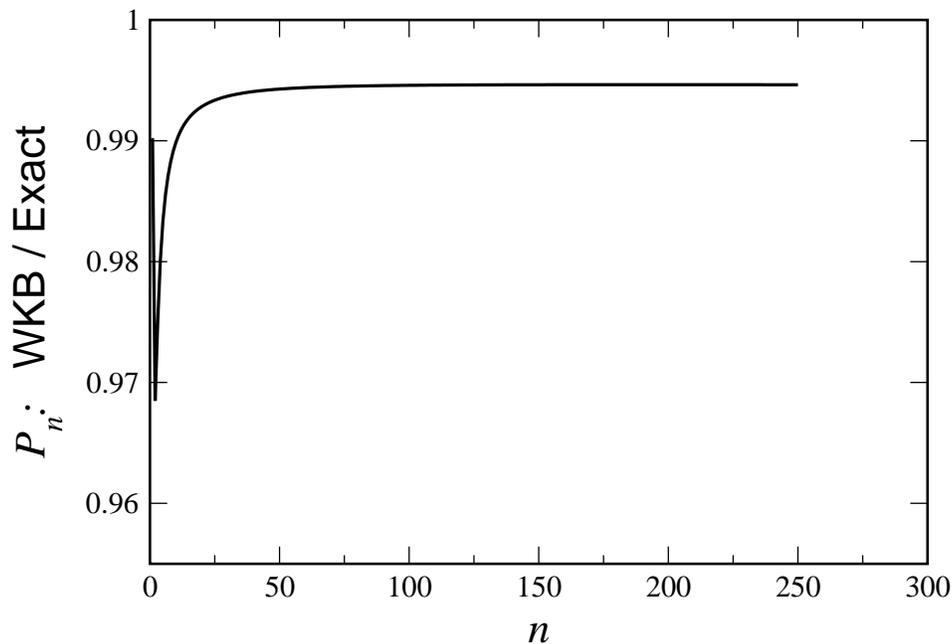}}
\caption{Ratio of the WKB approximation for $P_n$ to the exact result for the
basic model, Eq. (\ref{basic}) for $\beta=0.01$, $\alpha=1$.}
\label{npprat}
\end{figure}

With this result for $Q_n$ in hand, we can finish the matching procedure.  For $n \ll \alpha/\beta$, $\Lambda_n$ is large as we noted, and
\begin{equation}
Q_n\approx A \frac{\sqrt{2}\alpha}{\beta n}
\end{equation}
Comparing to the recursion relation results, we have
\begin{eqnarray}
A&=&P_2 \frac{\beta^2}{2\alpha^2}\nonumber\\
&=& P_1 \sqrt{\frac{\beta^{3}}{4\pi\alpha^3}}
\label{npa}
\end{eqnarray}
This fixes the ratio of $P_1$ to $P_2$, which is precisely that
which makes $\ln(P_n)$ a smooth function.   About the maximum,
$\Lambda_n \approx 1$, so that $Q_n\approx 4A/\sqrt{3}$, and
\begin{equation}
P_n \approx \frac{2}{\sqrt{3}}\left(\frac{\beta}{\alpha}\right)^2 P_2 e^{\Delta S} e^{-y^2/3}
\label{npfp}
\end{equation}
The sum over $P_n$ is dominated by this FP Gaussian, and so, replacing the sum by and integral, we get
\begin{equation}
P_{\it tot} \approx 2\sqrt{\pi} \left(\frac{\beta}{\alpha}\right)^{3/2} P_2 e^{\Delta S}
\end{equation}
The decay rate $\Gamma$ is then
\begin{equation}
\Gamma = \beta \frac{P_2}{P_{\it tot}} \approx \sqrt{\frac{\alpha^3}{4\pi\beta}}e^{-2\alpha(1-\ln(2))/\beta}
\label{npfinal}
\end{equation}
This result is plotted in Fig. \ref{np}, where we see the agreement
is excellent for small $\beta$.  As $\Gamma$ varies by so many
orders of magnitude over the scale of the graph, it is impossible to
see from this the role of the prefactor. In Fig. \ref{pref}, we plot
the ratio of $\Gamma$ to $e^{\Delta S}$ as a function of $\beta$. We
see the accuracy is quite good and gets better as $\beta$ gets
smaller, as expected. This prefactor  is, it should be noted, quite
different from the factor $\alpha$ conjectured in Ref. \cite{ek}. In Fig.
\ref{nppfig}, we present the graph of $P_n$ together with the WKB,
Fokker-Planck, and low/intermediate $n$ approximations.  We see that
the WKB approximation is excellent everywhere, whereas the other
approximations are more limited in their range of validity. In
particular, the Fokker-Planck results is a serious overestimate of
of the low $n$ probability, and an equally bad underestimate of the
large $n$ probability. As the WKB and exact results cannot be
distinguished on the scale of the figure, in Fig. \ref{npprat} we
present the ratio of the WKB to the exact result for $\alpha=1$,
$\beta=0.01$.  We see that the WKB approximation is good to the
expected few percent level, with it degrading slightly for very
small $n$.  Given that the WKB approximation is a large-$n$
approximation, that it does as well as it does at low-$n$ is a
undeserved present, and a consequence of the remarkable accuracy of
Stirling's formula down to $n=1$.

\begin{figure}[h]
\center{\includegraphics[width=0.7\textwidth]{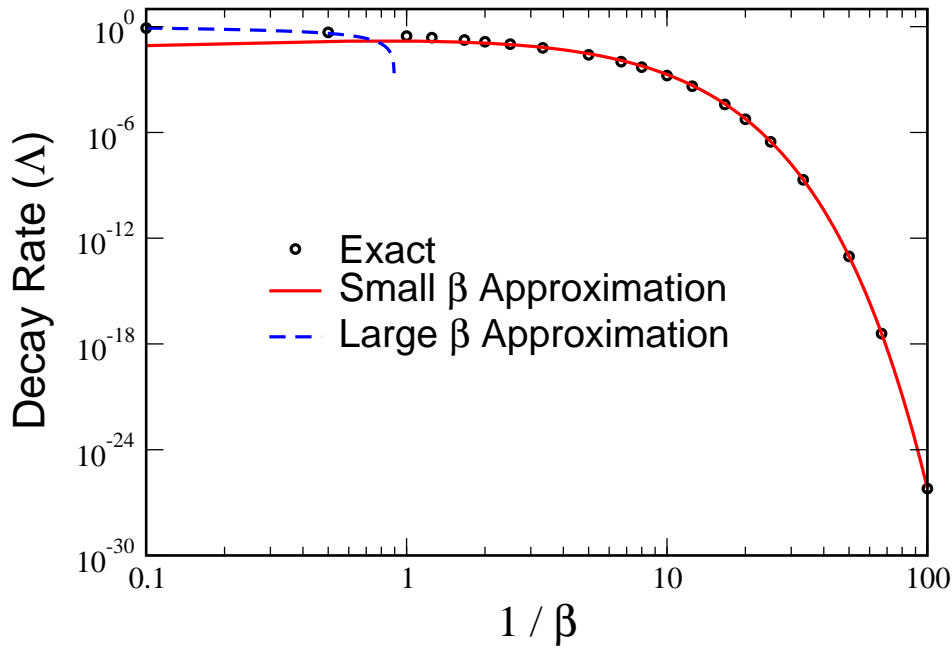}}
\caption{Decay rate vs. $1/\beta$ for the basic model, Eq. (\ref{basic}), with $\alpha=1$, together
with the large $\beta$ approximation, Eq. (\ref{nplargeb}), and the small $\beta$
approximation, Eq. (\ref{npfinal}).}
\label{np1}
\end{figure}

For completeness, we note that it is possible to derive a large
$\beta$ expansion of $\Gamma$ as well.  This is easily done by
truncating the master equation matrix and computing its determinant
as a power series in $\alpha/\beta$, then solving for $\Gamma$ order
by order.  One finds
\begin{equation}
\Gamma \approx \alpha - 2\frac{\alpha^2}{\beta}+\frac{10}{3}\frac{\alpha^3}{\beta^2}
-\frac{38}{9}\frac{\alpha^4}{\beta^3} + \frac{242}{135}\frac{\alpha^5}{\beta^4}
\label{nplargeb}
\end{equation}
It seems clear that this series has at best a finite radius of convergence.  Both large and small $\beta$ limits are compared to the exact results
in Fig. \ref{np1}.

\section{Parity-Preserving Model}

In this section we will discuss the case where birth is to "twins",
so that the even-odd parity of the number of particles is preserved.
Here, a series of mathematical "miracles" occur, which allow for a
simple calculation of the extinction rate (for the case of an
initial even number of particles) without recourse to the WKB
method. Furthermore, we present a solution of the probability
distribution to {\em all} orders in $\beta/\alpha$, so that the
corrections are exponentially small.  We also calculate the
extinction rate to all orders in perturbation theory and manage to
resum this divergent asymptotic series to obtain results correct to
within exponentially small terms.

The model,
\begin{eqnarray}
P&\stackrel{\alpha/2}{\to}& 3P\nonumber\\
P+P&\stackrel{\beta}{\to}& 0
\end{eqnarray}
conserves the even/odd parity of the number of particles, so if the system is initialized with
an odd number of particles it can never go extinct, instead reaching a steady-state.  If the system is initialized
with an even number of particles, on the other hand, the system can go extinct, with the survival probability
decaying again as $e^{-\Gamma t}$, with $\Gamma$ exponentially small for
small $\beta$. It should be noted that the mean-field equation for the model is the
exact same as that of the original model above.

We again start with the master equation, which now reads:
\begin{equation}
\dot{P}_n = -\Gamma P_n = \frac{\alpha}{2}\left[-nP_n + (n-2)P_{n-2}\right] + \frac{\beta}{2}\left[-n(n-1)P_n + (n+2)(n+1)P_{n+2}\right]
\end{equation}
As above, $\Gamma$ is exponentially small, and we may drop this term altogether.
We again tackle the master equation by exploiting the fast growth of the $P_n$'s for
not too large $n$.  This observation allows us to drop the second $\alpha$ term
and the first $\beta$ term, yielding the recursion relation
\begin{equation}
P_{n+2}=\left(\frac{\alpha}{\beta}\right)\frac{n}{(n+1)(n+2)}P_n
\end{equation}
which, up to a factor of 2, is the same as the approximate recursion relation we
previously encountered.  Now, however, the parity conservation implies that only
the even terms are nonzero, with the odd terms being exactly decoupled.
  The recursion relation has the solution
\begin{equation}
P_{2k}=\left(\frac{\alpha}{\beta}\right)^{k-1}\frac{(k-1)!2^{k}}{(2k)!}P_2
\label{solP}
\end{equation}

In principle, we should have to match this solution to the WKB solution, as we did
in the nonparity case.
The first miracle we encounter is that in fact the solution Eq. (\ref{solP}) is accurate throughout the Fokker-Planck region.  To see this, note that for $n=\alpha/\beta + y\sqrt{\alpha/\beta}$, $y \sim O(1)$,
the asymptotic expansion of $P_{n}$ is
\begin{equation}
 P_n \approx \sqrt{2} \left(\frac{\beta}{\alpha}\right)^2 e^{\alpha/2\beta} P_2 e^{-y^2/4}
\end{equation}
It is straightforward to verify that this is the solution of the Fokker-Planck equation:
\begin{equation}
 0= \alpha\left( \frac{d}{dy}(yP) + 2\frac{d^2}{dy^2}P\right)
\end{equation}
It should be noted for the record that while Eq. (\ref{solP}) is an accurate representation
of $P_n$ from $n=2$ till past the peak, the Fokker-Planck
Gaussian is again only valid in the peak region.

Thus, we can use Eq. (\ref{solP}) to calculate $P_{\it tot}\equiv \sum_k P_{2k}$.  We find
\begin{equation}
P_{\it tot}\approx P_2 \sum_{k=1}^{\infty} \left(\frac{\alpha}{\beta}\right)^{k-1}\frac{(k-1)!2^{k}}{(2k)!}={}_2F_2(1,1;\frac{3}{2},2;\frac{\alpha}{2\beta})P_2
\end{equation}
For the moment, what is important is the leading order asymptotics of this, which can be
calculated directly by applying Laplace's method to the sum.  This is equivalent to
integrating the Gaussian, and gives
\begin{equation}
P_{\it tot}\approx\sqrt{2\pi}\left(\frac{\beta}{\alpha}\right)^{3/2}e^{\alpha/2\beta}P_2
\end{equation}

 We are essentially done.  The rate of probability flux out to the absorbing state is $\beta P_2$, which equals
 $\Gamma P_{\it tot}$.  Thus,
 \begin{equation}
 \Gamma = \frac{\beta P_2}{P_{\it tot}} \approx \frac{\beta}{\sqrt{2\pi}\left(\frac{\beta}{\alpha}\right)^{3/2} e^{\alpha/2\beta}}=\sqrt{\frac{\alpha^3}{2\pi\beta}}e^{-\alpha/2\beta}
 \end{equation}
 Note again that this is much smaller than the naive Fokker-Planck answer, which is
 proportional to $e^{-\alpha/4\beta}$.

We can actually proceed to compute the corrections to this formula.  One source of
corrections is using the asymptotics of ${}_2F_2$.  The full asymptotics of ${}_2F_2$
for large argument are very beautiful:
\begin{equation}
{}_2F_2(1,1;\frac{3}{2},2;x)\approx \sqrt{\frac{\pi}{4x^3}}e^x \left(1 + \sum_{k=1}^\infty \frac{(2k-1)!!}{(2x)^k}
\right)
\end{equation}
This is obviously a divergent series, and alternatively represents a resummation of
the series.  This result is easily proven by substituting it in the Hypergeometric Differential
Equation.

The second source of the corrections is the corrections to the $P_n$'s due to the
terms we dropped in the master equation.  The structure here is also strikingly
beautiful.  If we denote our zeroth-order approximation of $P_{n}$ by $P_n^0$, we find
\begin{eqnarray}
P_4 &=& P_4^0\left(1 + \frac{\beta}{\alpha}\right)\nonumber\\
P_6 &=& P_6^0\left(1 + \frac{\beta}{\alpha}+ 3\left(\frac{\beta}{\alpha}\right)^2 \right)\nonumber\\
P_8 &=& P_8^0\left(1 + \frac{\beta}{\alpha}+ 3\left(\frac{\beta}{\alpha}\right)^2 + 15\left(\frac{\beta}{\alpha}\right)^3 \right)
\end{eqnarray}
The general trend is obvious:
\begin{equation}
P_{2k}=P_{2k}^0\left(1 + \sum_{m=1}^{k-1} (2m-1)!!\left(\frac{\beta}{\alpha}\right)^m\right)
\end{equation}
Plugging this into the master equation shows that this is an exact solution (for $\Gamma=0$, of course).  Now, up to exponentially small corrections, $P_{\it tot}$
is just multiplied by the correction factor:
\begin{eqnarray}
P_{\it tot}&=& P_{\it tot}^0 \left(1 + \sum_{m=1}^\infty (2m-1)!!\left(\frac{\beta}{\alpha}\right)^m\right)\nonumber\\
 &=& P_{\it tot}^0 \sqrt{\frac{\alpha^3}{2\pi\beta^3}}e^{-\alpha/2\beta}{}_2F_2(1,1;\frac{3}{2},2;\frac{\alpha}{2\beta})
\end{eqnarray}
The final result for $\Gamma$ is
\begin{equation}
 \Gamma \approx \frac{\sqrt{2\pi\beta^5/\alpha^3}}{\left[{}_2F_2(1,1;\frac{3}{2},2;\frac{\alpha}{2\beta})\right]^2}e^{\alpha/2\beta}
\label{final}
\end{equation}
Even though this answer is a resummation of the full asymptotic series, it is nevertheless not exact.  The correction
terms however are exponentially small, (relative to the exponentially small extinction rate) going like $e^{-\alpha/\beta}$.
This can be seen in the following graph, where we plot the relative error, comparing to an essentially exact numerical
calculation (using extended precision arithmetic in Maple).  Thus, for example, the error for $\alpha/\beta=100$ is six parts in $10^{21}$!
\begin{figure}
 \includegraphics[width=0.6\textwidth]{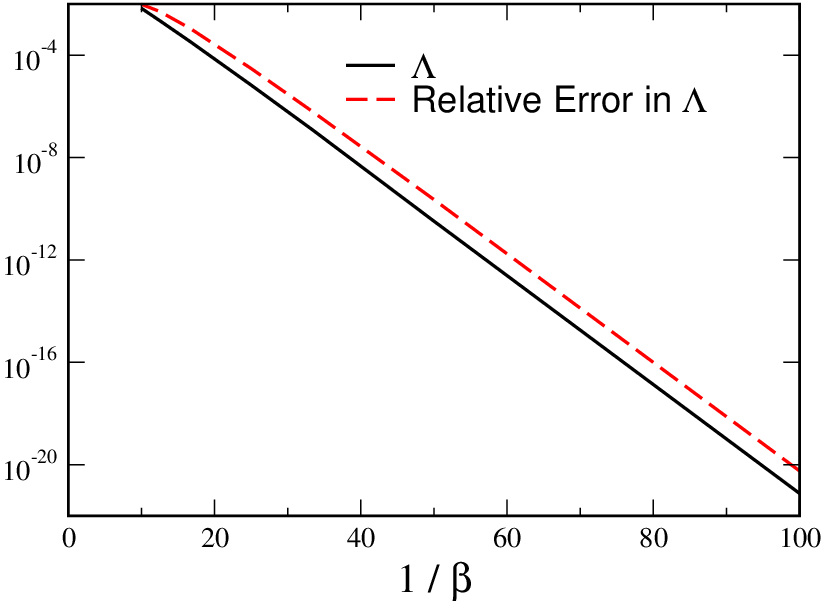}
\caption{Extinction rate $\Gamma$ and the relative error between the approximation, Eq. (\ref{final}), and the exact value
as a function of $1/\beta$ for $\alpha=1$.}
\label{relerr}
\end{figure}

For completeness, we also briefly write down the WKB solution. Firstly,
\begin{equation}
n(\Lambda)=\frac{\alpha}{\beta \Lambda^2}
\end{equation}
so that
\begin{equation}
\Delta S = \int_1^\infty \frac{n}{\Lambda}d\Lambda = \frac{\alpha}{2\beta}
\end{equation}
Also,
\begin{equation}
S_n = \frac{n}{2}\left[\ln(\alpha/\beta) + 1 - \ln(n)\right]
\end{equation}
The solution for $Q_n$ is the simple result
\begin{equation}
Q_n=A\Lambda^2
\end{equation}
All this can be seen to agree with our recursion relation solution.  In fact, it implies
that the recursion relation solution is valid everywhere, even past the FP regime.

\begin{figure}
\includegraphics[width=0.8\textwidth]{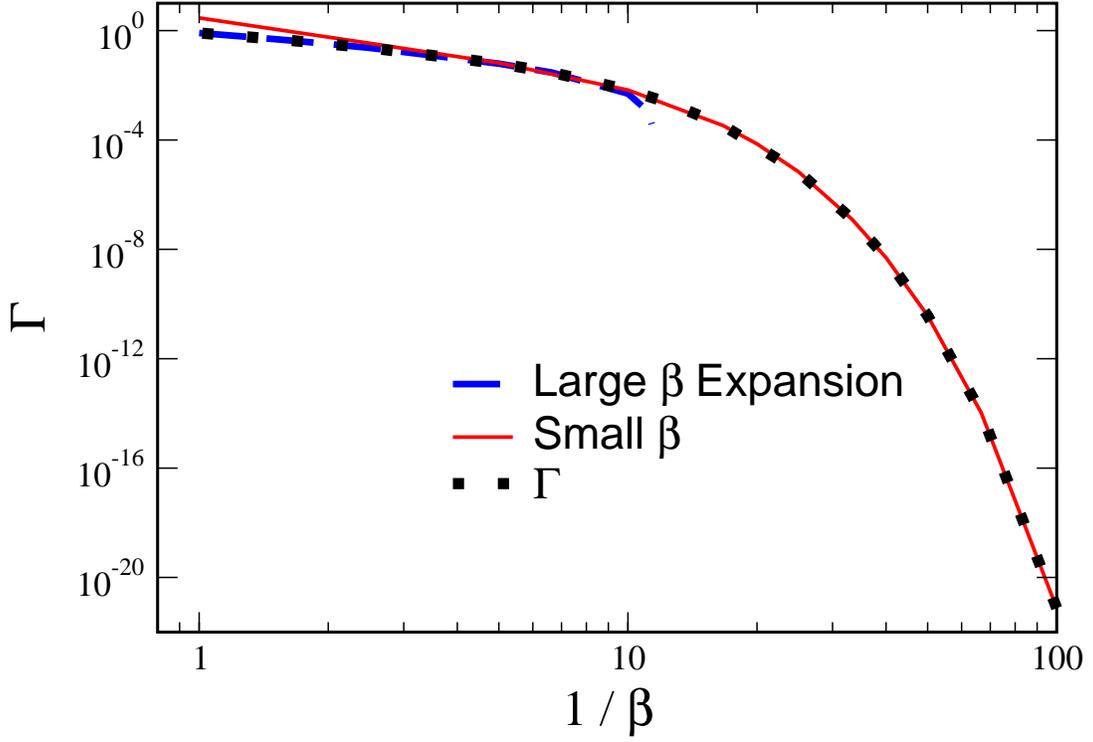}
\caption{Decay rate $\Gamma$ for the parity conserving model, compared to the
small $\beta$ result, Eq. (\ref{final}), and the large $\beta$ power-series expansion, Eq. (\ref{plarge}).  The agreement between these analytical expressions and the
exact results is so good in their respective spheres of validity, that the deviations
are only visible quite far outside these areas.}
\label{compar}
\end{figure}

One can again calculate the large $\beta$ limit of the decay rate.   The first few terms of this series are:
\begin{equation}
\Gamma = \beta - \frac{\alpha}{5}+\frac{12\alpha^2}{875\beta}-\frac{4\alpha^3}{21875\beta^2}-\frac{972\alpha^4}{58953125\beta^3}+\frac{24964\alpha^5}{95798828125\beta^4}
\label{plarge}
\end{equation}
It would appear likely that this series is actually convergent.  In any case, together with the
small $\beta/\alpha$ results above, they cover the entire range of parameters, as can be
seen in Fig. \ref{compar}.

\section{General Nonparity Model\label{gamsec}}

Let us  extend, now, our original, non-parity preserving, model to
include a third process, the spontaneous death of particles at a rate
$\gamma<\alpha$. Such a process appears naturally in many systems,
from populations of animals (where $\gamma$ is the death rate of an
individual) to the spread of a disease (where it correspond to a
recovery of an infected agent, like in the SIR model \cite{3}). We
present the physical optics solution for this case also, again
confirming it by comparison to the direct numerical solution.

Adding the  spontaneous decay of particles:
\begin{equation}
P\stackrel{\gamma}{\to} 0
\end{equation}
to our basic model, Eq. (\ref{basic}), changes both the mean field
and the fluctuations. At the mean-field level this is equivalent to
a simple change of the  effective growth rate, $\alpha$ of
$\alpha_{\it eff}=\alpha-\gamma$, but this scaling is not true
anymore if the fluctuations are taken into account. The master
equation now reads:
\begin{equation}
\dot{P}_n = \alpha\left[-nP_n + (n-1)P_{n-1}\right] +
\frac{\beta}{2}\left[-n(n-1)P_n + (n+2)(n+1)P_{n+2}\right] +
\gamma\left[-nP_n + (n+1)P_{n+1}\right].
\end{equation}
We start this time with the WKB solution.  As before, the WKB ansatz
yields an equation for $\Lambda_n$, namely
\begin{eqnarray}
0&=&\left(\frac{1}{2\Lambda_n}\right)\left[2\alpha\left(-\Lambda_n+1\right) + \beta n\left(-\Lambda_n + \Lambda_n^3\right) + 2\gamma\left(-\Lambda_n + \Lambda_n^2\right)\right]\nonumber\\
&=& \left(\frac{\Lambda_n-1}{2\Lambda_n}\right)\left[-2\alpha+\beta n \Lambda_n(\Lambda_n +1) +2\gamma\Lambda_n\right]
\end{eqnarray}
with the solution
\begin{equation}
\Lambda_n=\frac{\sqrt{(\beta n + 2\gamma)^2 + 8\alpha\beta n}-2\gamma-\beta n}{2\beta n}
\end{equation}
Now, $\Lambda$ approaches the finite limit, $\alpha/\gamma$ as $n$ goes to 0, in
contrast to the previous cases.  Thus, we cannot solve the low-$n$ recursion relation
by assuming that the $P_n$ are increasing very rapidly.  Rather, now $\beta$ is irrelevant
for low $n$, and we have to solve the $\beta=0$ recursion.  This is readily solved, for
example by generating function techniques, and yields
\begin{equation}
P_n=\frac{\gamma}{n(\alpha-\gamma)}\left[\left(\frac{\alpha}{\gamma}\right)^n-1\right]P_1
\label{lowng}
\end{equation}
Indeed, for large $n$ $P_n$ grows geometrically, with the ratio $\alpha/\gamma$,
agreeing with the small $n$ WKB.

We can now proceed with the remainder of the WKB procedure.  As before, it is more
convenient to work with $n(\Lambda)$, given by
\begin{equation}
n(\Lambda)=\left(\frac{2}{\beta}\right)\frac{\alpha-\gamma\Lambda}{\Lambda(\Lambda+1)}
\end{equation}
We see that in the mean-field regime, $\Lambda\approx1$, the entire $\gamma$ dependence
is through $\alpha_{\it eff}$.  Away from this limit, however, the situation is more complicated.

\begin{figure}
\center{\includegraphics[width=0.7\textwidth]{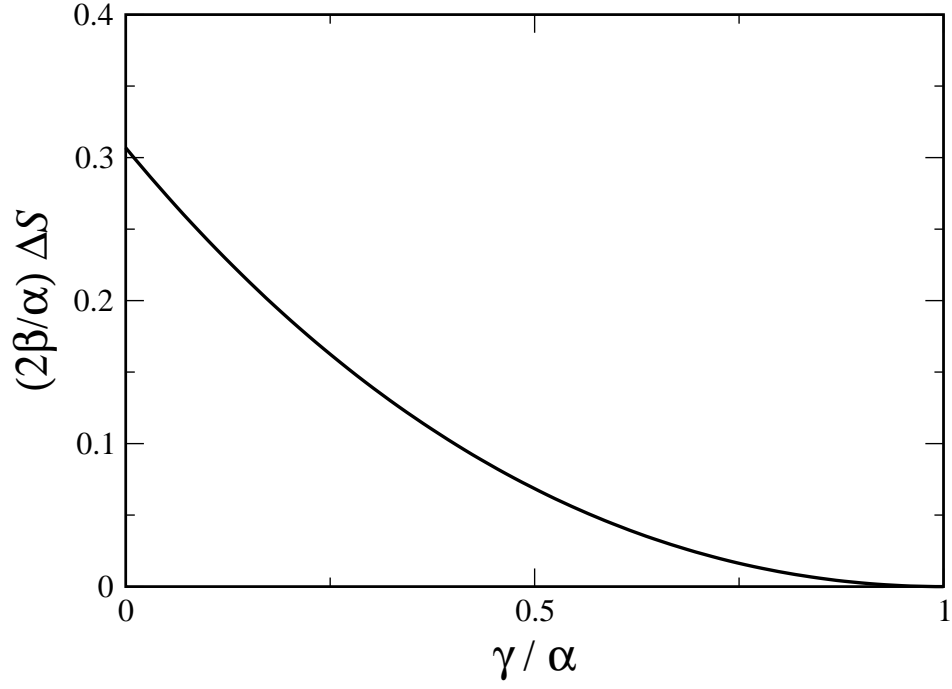}}
\caption{$2\beta\Delta S/\alpha$ as a function of $\gamma$, for fixed $\alpha$.}
\label{fixalf}
\end{figure}

\begin{figure}[b]
\center{\includegraphics[width=0.7\textwidth]{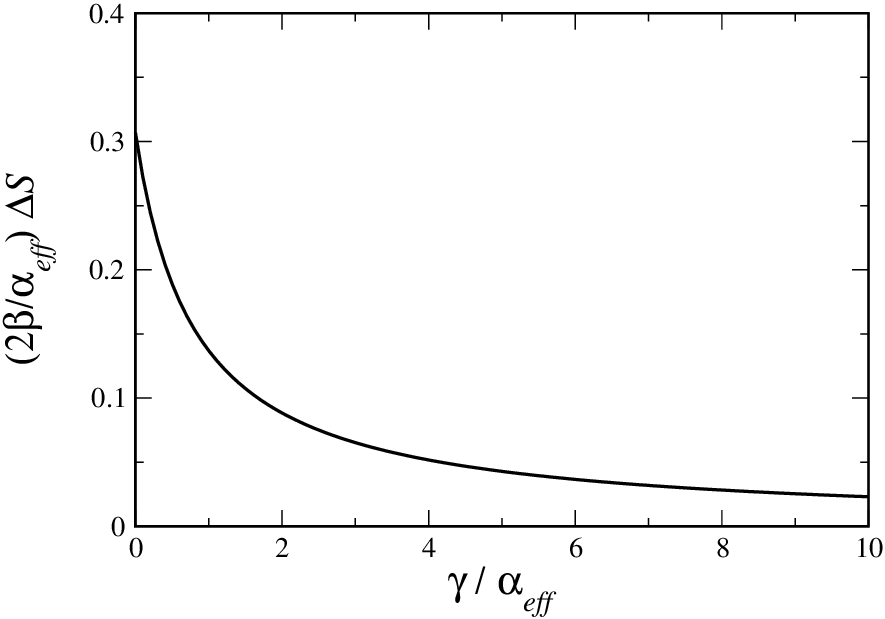}}
\caption{$2\beta\Delta S/\alpha_{\it eff}$ as a function of $\gamma$, for fixed $\alpha_{\it eff}$.}
\label{fixeff}
\end{figure}

Now, as before, $S_n$ is given by
\begin{eqnarray}
S_n&=& S_0 - \left[n(\Lambda)\ln(\Lambda)\rule{0cm}{12pt}\right]_{\Lambda}^{\alpha/\gamma} + \int_\Lambda^{\alpha/\gamma} \frac{n(\Lambda}{\Lambda}d\Lambda\nonumber\\
&=&S_0 + n\ln(\Lambda) +\frac{2(\alpha-\gamma\Lambda)}{\beta\Lambda} - \frac{2(\alpha+\gamma)}{\beta}\left(\ln\left(\frac{\Lambda+1}{\Lambda}\right)-\ln\left(\frac{\alpha+\gamma}{\alpha}\right)\right)
\end{eqnarray}
In particular,
\begin{equation}
\Delta S=\frac{2(\alpha-\gamma)}{\beta}+\frac{2(\alpha+\gamma)}{\beta}\ln\left(\frac{\alpha+\gamma}{2\alpha}\right)
\label{npdels}
\end{equation}
We exhibit $\Delta S$ as a function of $\gamma$ for fixed $\alpha$ in Fig. \ref{fixalf}.  We
see that $\Delta S$ decreases with increasing $\gamma$, vanishing quadratically as
$\gamma$ approaches the threshold value of $\alpha$.  Also interesting is the dependence of $\Delta S$ as a function of $\gamma$, for fixed $\alpha_{\it eff}$.  This
is show in Fig. \ref{fixeff}.  We see that as $\gamma$ increases, at fixed $\alpha_{\it eff}$,  $\Delta S$ decreases, leading to a faster decay rate due to increased fluctuations. For large $\gamma$,
in fact, $\Delta S$ vanishes as  $\alpha_{\it eff}/(2\beta \gamma)$.

We now are in a position to continue to the calculation of $Q_n=e^{S_1(n)}$.  The equation is
\begin{eqnarray}
 0&=&\alpha\left[-n + \frac{n}{\Lambda_n}\left(1 - S_1' + \frac{1}{2}S_0''  - \frac{1}{n}\right)\right] \nonumber\\
 &\ &{} + \frac{\beta}{2}\left[-n^2 + n + n^2\Lambda_n^2\left(1 + 2S_1'+2S_0''+\frac{3}{n}\right)\right] \nonumber\\
 &\ &{} +\gamma\left[-n + n\Lambda_n\left(1 + S_1'+\frac{1}{2}S_0'' + \frac{1}{n}\right)\right]
 \end{eqnarray}
 which simplifies to, using $S''=1/(\Lambda n'(\Lambda))$,
 \begin{equation}
\left [\frac{\alpha}{\Lambda}-\beta n\Lambda^2-\gamma\Lambda\right]\frac{d}{d\Lambda}S_1=\frac{1}{2\Lambda}\left[\frac{\alpha}{\Lambda}+2\beta n\Lambda^2 + \gamma \Lambda\right] + \frac{n'(\Lambda)}{2n}\left[-\frac{2\alpha}{\Lambda}+\beta n + 3\beta n\Lambda^2 + 2\gamma\Lambda\right]
 \end{equation}
 This has the solution
 \begin{equation}
 Q_n=A\frac{\sqrt{\alpha^3\Lambda_n}(\Lambda_n+1)^2}{(\alpha-\gamma\Lambda_n)\sqrt{\alpha(2\Lambda_n+1)-\gamma\Lambda_n^2}}
 \end{equation}
 where we have inserted the factor $\sqrt{\alpha^3}$ so the definition of $A$ reduces to that used in the $\gamma=0$ case.

\begin{figure}
\center{\includegraphics[width=0.7\textwidth]{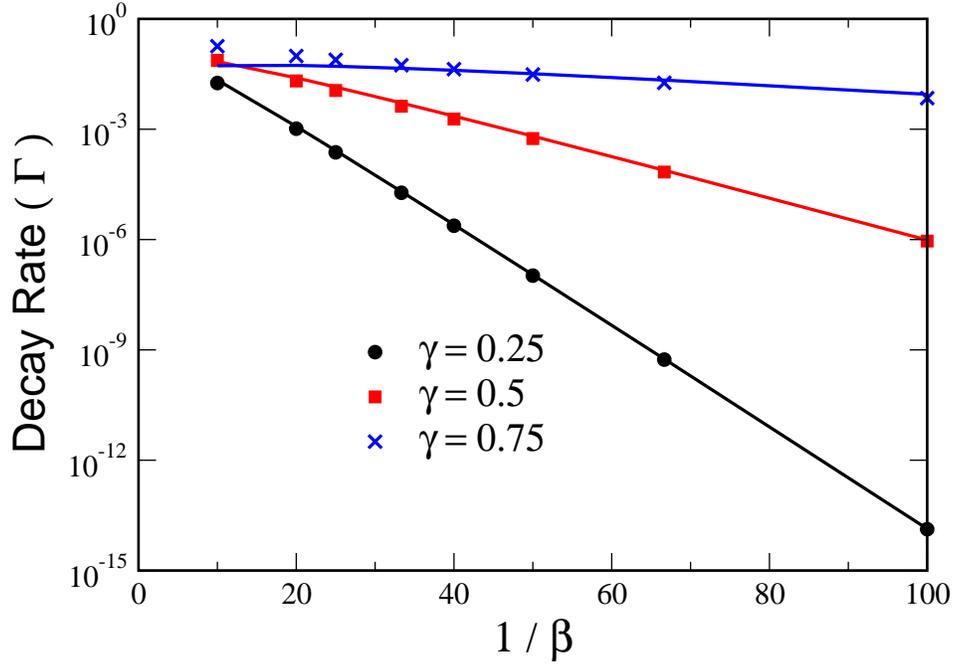}}
\caption{The calculated decay rate $\Gamma$ as a function of $1/\beta$, Eq. (\ref{gamgam}) for $\alpha=1$, $\gamma=0.25$, $0.5$, and $0.75$ (solid line) together with the exact numerical results (markers).}
\label{gamfig}
\end{figure}

 The last step is to match to the low-$n$ recursion relation solution. For $\beta n \ll 1$,
 the WKB solution reduces to
 \begin{equation}
 P_n\approx \frac{2A\sqrt{\alpha(\alpha+\gamma)}}{\beta n}\left(\frac{\alpha}{\gamma}\right)^n
 \end{equation}
 Comparing this to Eq. (\ref{lowng}), we get
 \begin{equation}
 A=\frac{\gamma\beta}{2(\alpha-\gamma)\sqrt{\alpha(\alpha+\gamma)}} P_1
 \end{equation}
 We now need to calculate $P_{\it tot}$.  Near the stable point,  $n=(\alpha-\gamma)/\beta$,
 \begin{equation}
 P_n=4\sqrt{\frac{\alpha^3}{3\alpha-\gamma}}\frac{A}{\alpha-\gamma}e^{\Delta S} e^{-\frac{\beta}{3\alpha-\gamma}\left(n-\frac{\alpha-\gamma}{\beta}\right)^2}
 \end{equation}
 which gives
 \begin{eqnarray}
 P_{\it tot}&=&4\sqrt{\frac{\pi\alpha^3}{\beta}}\frac{A}{\alpha-\gamma}e^{\Delta S}\nonumber\\
 &=& \sqrt{\frac{\pi\beta}{\alpha+\gamma}}\frac{2\alpha\gamma}{(\alpha-\gamma)^2}P_1e^{\Delta S}
 \end{eqnarray}
 The total probability flux out of the system is $\gamma P_1 + \beta P_2$.  Since $P_2$ is of order 1 relative to $P_1$, the $\beta$ contribution to the flux is negligible, and so
 \begin{equation}
 \Gamma = \frac{\gamma P_1}{P_{\it tot}} = \sqrt{\frac{\alpha+\gamma}{\pi \beta}}\frac{(\alpha-\gamma)^2}{2\alpha}e^{-\Delta S}
 \label{gamgam}
 \end{equation}
 This result is tested against the exact numerical answer in Fig. \ref{gamfig}.  The results are again quite good, with the quality decreasing as $\gamma$ approaches $\alpha$ (for fixed $\beta$) as expected due to the decreasing equilibrium   number of particles. The astute reader will note that our expression for $\Gamma$ reduces in the
 $\gamma\to 0$ limit to our $\gamma=0$ result above, even though some intermediate
 expressions (for example, the probability flux) do not correspond.  We also note that
 the maximum value of $\Lambda$ is $\alpha/\gamma$, which approaches 1 as
 $\gamma$ approaches the threshold value of $\alpha$.  Thus, near threshold,
 the Fokker-Planck equation and the WKB treatment coincide.  Of course, this
 solution still has to be matched to the low/intermediate-$n$ solution to obtain
 the correct prefactor.

 It should also be noted that Doering, et al.,\cite{b} investigated a class of models wherein
 all transitions are single-particle transitions, as opposed to the models
 investigated herein, which include a 2 particle annihilation
 process. This entire class of models can also be easily treated via our
 WKB method, and yields identical results to those of Doering, et al.  Writing the birth term in the master equation
 by $\alpha_n$ and the death term by $\gamma_n$, the WKB solution can be written
 down for a very broad class of models where $\alpha_n=(1/\beta)\tilde{\alpha}(\beta n)$ and $\gamma_n=(1/\beta)\tilde{\beta}(n)$, and $\tilde{\alpha}$, $\tilde{\beta}$ are smooth
 functions. In this case, the system admits a macroscopic metastable state with a large number of
 particles for small $\beta$.  In particular, we find
 \begin{equation}
 \Lambda_n = \frac{\alpha_n}{\gamma_n}
 \end{equation}
 so that
 \begin{equation}
 S_n = \int_0^n \ln(\Lambda_n) dn
 \end{equation}
 The $Q_n$ factor is:
 \begin{equation}
 Q_n=\frac{A}{\sqrt{\alpha_n\gamma_n}}
 \end{equation}
 which, upon matching to the low/intermediate-$n$ result gives (up to an obvious
 typographical error) the result in Doering, et al. The advantage of the WKB method
 is that it generalizes to the multi-particle transition case.

 \section{Threshold case and the lifetime of a neutral mutation}
 In that section let us consider the threshold case,
 $\alpha=\gamma$. This case corresponds to the dynamics of a neutral
 mutation and has recently become the focus of extensive research,
 mainly in connection with Hubbell's unified neutral theory of
 biodiversity and biogeography.\cite{4}
 Here, as we already
 saw for the near-threshold case, the $P_n$'s are
 smooth and allow for a Fokker-Planck treatment.  However, in this case the
 decay rate $\Gamma$ is not exponentially small, and so cannot be ignored.
 First, let us consider what happens in the absence of $\beta$.  This
problem was worked out by Pechenik and Levine.\cite{PL}  They find that the mean number of particles is conserved and the
variance grows linearly in time.  Furthermore, the system exhibits a power-law ($1/t$) convergence to the empty state,
and not an exponential dependence.  This is indicative of a scale invariance in the problem.  The $\beta$ term serves
to break this scale invariance, and gives a well-defined scale for the number of particles (for those replicas which still
survive).  The original Fokker-Planck equation (ignoring boundary terms) was
\begin{equation}
 \frac{\partial P}{\partial t} = \alpha \frac{\partial^2}{\partial n^2} (nP)
\end{equation}
The $\beta$ process introduces two new terms into the FP equation, arising from the Taylor expansion of
$\frac{1}{2}\beta\left(-n(n-1)P(n) + (n+2)(n+1)P(n+2)\right)$ in the master equation.  The first of these terms is a drift term, $\beta\frac{\partial}{\partial n} (n^2 P)$.  This is responsible for the term in the mean-field equation.  The second term is a diffusion term.  If we
assume $\beta$ small, then the additional diffusion induced by $\beta$ can be ignored, and we are left with only the drift term.  The new FP equation now reads
\begin{equation}
 \frac{\partial P}{\partial t} = -\Lambda P = \alpha \frac{\partial^2}{\partial n^2} (nP) + \beta\frac{\partial}{\partial n} (n^2 P)
\end{equation}
We have assumed the time dependence is exponential, and are looking for the smallest eigenvalue $\Lambda$.  We
shall see that $\Lambda$ vanishes in the $\beta\to 0$ limit, consistent with the power-law behavior found by P-L.

\begin{figure}
\center{\includegraphics[width=0.7\textwidth]{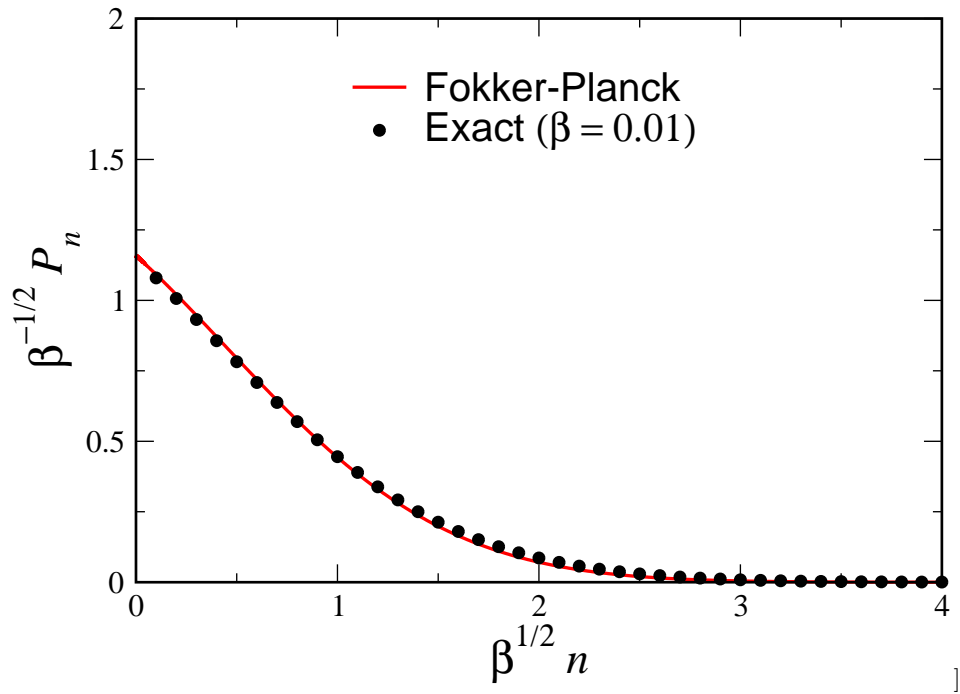}]}
\caption{Rescaled probability distribution $\beta^{-1/2}P_n$ as a function of
$\beta^{1/2} n$ for the case $\alpha=\gamma=1$, $\beta=0.01$, together with
the numerical solution of the Fokker-Planck equation.}
\label{threshp}
\end{figure}

It is useful to transform the FP equation to Shroedinger form.  The first step in this
process is to change variables to $x\equiv \sqrt{n}$.  This yields the equation
\begin{equation}
-\alpha P'' - (\frac{3\alpha}{x} + 2 \beta x^3)P' -8\beta x^2 P = 4\Gamma P
\end{equation}
The next step is a similarity transformation to eliminate the first derivative
\begin{equation}
P\equiv x^{-3/2}e^{-\beta x^4/4\alpha} Q
\end{equation}
yielding the Schroedinger equation
\begin{equation}
-Q'' + \left(\frac{3}{4x^2} - \frac{2\beta x^2}{\alpha} + \frac{\beta^2 x^6}{\alpha^2}\right) Q = \frac{4\Gamma}{\alpha}Q
\label{threshfpq}
\end{equation}
Clearly, in the absence of $\beta$ there is no bound state. Rather there is a continuum that starts at zero.  The potential with $\beta$ has a single minimum, which is negative.
Nevertheless, the ground state energy is positive, yielding a decay rate.  The scaling
of the decay rate is clear; by rescaling $y\equiv (\beta/\alpha)^{1/4} x$, $\beta$ and
$\alpha$ disappear from the equation, with the decay rate scaling as $\sqrt{\alpha \beta}$.  As advertised, we verify the vanishing of $\Lambda$ with $\beta$.  The presence of
$\beta$ has set the scale of (the surviving) $n$'s, namely $\sqrt{\alpha/\beta}$, which
is large for small $\beta$.  Lastly, the $1/x^2$ nature of the $\alpha$ potential is
clearly a result of the scale-free nature of the $\beta=0$ problem. To get the prefactor
multiplying the $\sqrt{\beta}$, we have to numerically solve the rescaled Shroedinger
equation, (i.e., Eq. (\ref{threshfpq}) with $\alpha=\beta=1$) yielding the result
\begin{equation}
\Gamma =1.111\sqrt{\alpha\beta}
\end{equation}
The resulting scaled $P_n$ is shown in Fig. \ref{threshp}, together with the rescaled
exact numerical solution of the master equation for $\beta=0.01$.  We see that
$P_n$ is strongly peaked at the origin, corresponding to the zero particle mean-field
solution.

The scaling with $\beta$ we obtained is the same as found by Doering, et al.  However, as opposed
to the other cases examined herein, the prefactor is different as now the mean time to extinction is not simply
the inverse of the decay rate.  This is due to the fact that all the eigenvalues of the
master equation scale as $\sqrt{\beta}$, whereas the other cases exhibited a single
exponentially small eigenvalue.

\section{Conclusions}
We have solved for the fluctuation-induced extinction rates of various models exhibiting  a
macroscopic metastable state. Our primary methodology is use the WKB method for difference
equations to directly solve the master equation.  This WKB solution then
has to be matched to the low-$n$ $P_n$'s, since the WKB method is a large $n$
approximation.  This technique is quite
general and straightforward to implement, and produces quite accurate results as long
as there are not too few particles in the metastable state.
It reproduces the Doering, et al. results for the case of general one-particle transitions and generalizes to higher-order
transitions.  We have also shown the unique mathematical properties of the even/odd
parity conserving model, where we are able to generate the full asymptotic expansion
for the decay rate to all orders, and even to resum this divergent asymptotic series.

One interesting point which arises from this analysis is the sensitivity of the decay rate to the exact form of the microscopic dynamics.  This is apparent in the very different dominant $e^{\Delta S}$ for the case of the parity and non-parity cases.  This is also
the case when one compares the logistic model with spontaneous decay studied in
Section \ref{gamsec}, to the same model where the collision process $P+P \to 0$ is
replaced by $P+P\to P$ at twice the rate.  Even at the level of the Fokker-Planck
dynamics valid near the metastable state, the widths of the distributions are different,
with the variance $(3\alpha-\gamma)/2\beta$ being replaced by $\alpha/\beta$.
The values of $\Delta S$ in the two cases are very different, where Eq.  (\ref{npdels})
for the two-particle annihilation should be compared to
\begin{equation}
\Delta S = \int_0^{(\alpha-\gamma)/\beta} \ln\left(\frac{\alpha}{\gamma + \beta n}\right)
= \frac{\alpha-\gamma}{\beta} - \frac{\gamma}{\beta}\ln\left(\frac{\alpha}{\gamma}\right)
\end{equation}
Thus, while the two expressions agree near threshold, where the Fokker-Planck description is sufficient, as $\gamma$ approaches 0, $\Delta S\to 2\alpha/\beta(1-\ln(2))$  for the two-particle
annihilation case, as we saw in Section 1, whereas $\Delta S \to \alpha/\beta$, almost
twice as large,
for small $\gamma$ in the $P+P\to P$ case.  Parenthetically, it should be noted
that the small $\gamma$ limit is very singular in this latter case, as naively
$\Gamma$ seems to diverge as $\gamma^{-1/2}$ for small $\gamma\ll \beta$.
In general, for small $\beta$ and $\gamma$, the decay rate can be shown to be given by
\begin{equation}
\Gamma \approx \left(\frac{\alpha}{\beta}\right)^{\gamma/\beta} \frac{a}{\Gamma(\gamma/\beta)} e^{-\alpha/\beta}
\end{equation}
as so indeed vanishes as $\gamma\to 0$, since there is no extinction in this case.
 In situations where
the Fokker-Planck equation is valid all the way down to $n=0$, as we
saw was the case at threshold, at least the problem is parameterized
by only two parameters, the center and width of the Gaussian.
However, in general, the entire function $n(\Lambda)$ is involved in
the calculation of $\Delta S$.  This should have important
implications for the study of extinctions in the ecological
community, for example, where reliable microscopic models are
difficult if not impossible to obtain.

\acknowledgments{The authors thank B. Meerson and M. Assaf for sharing their work prior to publication.  The work of DAK is supported in part by the Israel Science Foundation.  The work of NMS is supported in part by the EU 6th framework CO3 pathfinder.}

\end{document}